\documentclass[10pt,conference]{IEEEtran}
\usepackage{hhline}
\usepackage{multirow}
\usepackage{multicol}
\usepackage{xspace}
\usepackage{fontawesome}
\usepackage{csquotes}
\usepackage[dvipsnames]{xcolor}
\usepackage[rightmargin=0.5pt,leftmargin=0.5pt]{quoting}
\usepackage{caption}
\usepackage{subcaption}
\usepackage[most]{tcolorbox}
\usepackage{amsmath}

\usepackage{amssymb}
\usepackage[breaklinks, hidelinks]{hyperref}
\usepackage[export]{adjustbox}

\usepackage{graphicx}
\usepackage{adjustbox}
\usepackage{balance}
\graphicspath{ {./figures/} }
\usepackage[most]{tcolorbox}
\tcbset{%
    bugreportstyle/.style={
        enhanced,
	    theorem name,        
        colback=white,
        colframe=black,
        fonttitle=\itshape,
        colbacktitle=white,
        coltitle=black,
        boxrule=1pt,
        boxsep=0pt,
        left=4pt,
        right=4pt,
        before skip=4pt,
        after skip=0pt,
        separator sign none,
        attach boxed title to top left={%
            yshift*=-\tcboxedtitleheight/2, 
            xshift=5mm},
        boxed title style={colframe=black}
    }
}
\newcommand\sbullet[1][.5]{\mathbin{\vcenter{\hbox{\scalebox{#1}{$\bullet$}}}}}

\newtcbtheorem{bugreport}{}{bugreportstyle}{}

\include{macro}

\begin{document}

\title{An Empirical Investigation into the Reproduction of Bug Reports for Android Apps}

\author{
	\IEEEauthorblockN{Jack Johnson$^{*}$, Junayed Mahmud$^{\dagger}$, Tyler Wendland$^{*}$, Kevin Moran$^{\dagger}$, Julia Rubin$^{\ddagger}$, Mattia Fazzini$^{*}$}
	\IEEEauthorblockA{
		\textit{$^{*}$University of Minnesota, MN, USA}; \href{mailto:}{joh19267@umn.edu}, \href{mailto:}{wendl155@umn.edu}, \href{mailto:}{mfazzini@umn.edu}\\
		\textit{$^{\dagger}$George Mason University, VA, USA}; \href{mailto:}{jmahmud@gmu.edu}, \href{mailto:}{kpmoran@gmu.edu}\\
		\textit{$^{\ddagger}$University of British Columbia, BC, Canada}; \href{mailto:}{mjulia@ece.ubc.ca}\\
	}
}

\maketitle

\begin{abstract}

One of the key tasks related to ensuring mobile app quality is the reporting, management, and resolution of bug reports. As such, researchers have committed considerable resources toward automating various tasks of the bug management process for mobile apps, such as reproduction and triaging. However, the success of these automated approaches is largely dictated by the characteristics and properties of the bug reports they operate upon. As such, understanding mobile app bug reports is imperative to drive the continued advancement of report management techniques. While prior studies have examined high-level statistics of large sets of reports, we currently lack an in-depth investigation of how the information typically reported in mobile app issue trackers relates to the specific details generally required to reproduce the underlying failures.

In this paper, we perform an in-depth analysis of 180 reproducible bug reports systematically mined from Android apps on GitHub and investigate how the information contained in the reports relates to the task of reproducing the described bugs. In our analysis, we focus on three pieces of information: the environment needed to reproduce the bug report, the steps to reproduce (S2Rs), and the observed behavior. Focusing on this information, we characterize failure types, identify the modality used to report the information, and characterize the quality of the information within the reports. We find that bugs are reported in a multi-modal fashion, the environment is not always provided, and S2Rs often contain missing or non-specific enough information. These findings carry with them important implications on automated bug reproduction techniques as well as automated bug report management approaches more generally. 
\end{abstract}

\section{Introduction}
\label{sec:intro}

The importance of the quality of mobile applications (colloquially referred to as apps) has grown in recent years as smartphones and tablets have become deeply integrated into users' daily lives. Once an application has been released to users, its quality is 
largely ensured by continuing maintenance activities, which have been shown to consume considerable amounts of engineering effort~\cite{2008_tassey_testing}. 
These important maintenance activities are typically centered around \emph{bug report management} and include activities related to understanding, reproducing, and resolving bug reports.

A number of unique development constraints related to mobile apps, such as 
pressure for frequent releases~\cite{Hu:ESYS14,Jones:2014}, 
the need to cope with constantly evolving platform APIs~\cite{2012_han_wcre_fragmentation,android-fragmentation}, 
a large volume of user feedback~\cite{Ciurumelea:SANER'17,DiSorbo:FSE'16,Palomba:ICSE17,Palomba:JSS'18,Palomba:ICSME15}, and testing challenges~\cite{Choudhary2015} complicate the bug report management process. 
Software engineering researchers have recognized these domain-specific challenges and have worked toward providing automated solutions across several bug report management activities for mobile apps, including
bug report quality assessment~\cite{Chaparro2019}, reproduction~\cite{Fazzini2018,Zhao2019}, triaging~\cite{Xia2017}, and bug localization~\cite{2020_issta_pradel_scaffle,zhang2019commit}.

One common thread among these various automated solutions is that they operate directly upon the information contained within bug reports and, as such, are directly affected by the characteristics and quality of various report components, such as environmental information (e.g., device, software version), reproduction steps (S2Rs), and observed behavior (OB). Thus, researchers and practitioners require a solid empirical foundation that delineates common characteristics of mobile app bug reports to build effective automated techniques. 

In prior work, researchers have examined high-level statistics (\eg number and type of report, fix rates, fix time) of large sets of bug reports. For example, Battacharya \etal~\cite{Bhattacharya2013} performed an empirical study on bugs submitted to the Android platform on 24 widely-used open source apps. Others have compared high-level bug characteristics between mobile apps and desktop apps~\cite{Zhou2015}. However, to the best of our knowledge, no study has yet provided an in-depth characterization of how the information contained in mobile bug reports might impact the task of bug reproduction. One likely reason that past studies have not examined this relation is that as it requires \textit{manually reproducing} real bug reports, which is a time-consuming and difficult task. Despite the difficulty of this analysis, understanding this information is critical as both developers and automated bug analysis techniques may need to (i) understand the type of reported failure, (ii) understand multiple modalities of information, such as text, images, or screen-recordings, and (iii) identify or infer information that is either vague or missing from the reports. In short,  empirically analyzing both the \textit{characteristics} and \textit{quality} of the information reported in mobile app bugs is critical for both the practical and scientific advancement bug report management for mobile apps.

In this paper, we conduct and in-depth characterization of reproducible bug reports for Android apps. 
To this end, we significantly extend \textsc{AndroR2}~\cite{Wendland2021} -- 
a dataset of reproducible bug reports for Android apps which contains bugs representing a range of failure types.
We augmented the dataset with additional, manually verified and fully reproduced bug reports from open source Android apps
hosted on GitHub~\cite{2021_github} and available on the Google Play store~\cite{2021_google_play}, 
obtaining a dataset of \brcount bug reports. In this work, we focus on bug reports for Android apps as Android is the most widely used operating system for mobile apps~\cite{StatCounter}.
To the best of our knowledge, ours is the largest dataset of 
(i) fully reproduced bug reports for Android apps, which 
(ii) contains both user-submitted and developer-submitted reports, and 
(iii) in contrast to related work, focuses on different types of failures beyond app crashes. 
Given this dataset, we focused our in-depth analysis on three sources of information: 
the description of the environment needed to reproduce the bug report, 
the steps to reproduce, and the observed behavior.

Leveraging the fact that our studied reports are considered fully reproducible, we perform an in-depth analysis of both the report \textit{characteristics}---including the failure types and modalities of reported information---and the \textit{quality} of reported information.  In relation to the quality of reported information, we focus on three aspects: the types and prevalence of missing information, whether report discussion threads contain helpful information for reproducing the reports, and the specificity of reported information (which investigates whether reported information can be directly used for reproducing the reports). Although these aspects are only some of ones that describe the quality of reported information, we believe that the analysis of these aspects provides useful insights into the reproduction of bug reports and hence focus on them.

Our analysis shows that 
(i) reported failures can be grouped into four types, three of which are not yet considered by existing automated reproduction techniques, 
(ii) different information modalities are used to report the details related to the environment, steps to reproduce, and observed behavior,
(iii) a large number of reports (74\%) have at least one step to reproduce that requires multiple operations in the app 
indicating that the information provided for the step is not always specific enough,
(iv) the great majority of reports (92\%) have at least one missing reproduction step, illustrating that the operations
required to reproduce the reports must often be inferred,
and (v) bug report discussions can, in some cases (19\%), provide additional information useful for the reproduction of the reports.
Finally, we discuss implications of our findings, which can help guide future research on automated reproduction of bug reports and, more generally, bug report management activities.

In summary, the main contributions of this paper are:

\begin{itemize}
	\item{A large set of 180 manually mined and reproduced bug reports for Android apps that contains user- and developer-submitted bug reports of multiple failure types.} 
	\item{A study that examines bug characteristics and information quality in reproducible mobile app bug reports. This advances upon prior studies which do not manually verify and collect reproducible bug reports.}
	\item{A discussion on the implications of our findings, which illustrates the need for future research on non-crashing oracles, multi-modal understanding of report information, mocking environments, and missing and non-specific reproduction steps.}
	\item{A replication package~\cite{appendix} that contains our dataset of bug reports, data analysis reports, and scripts to perform the study analyses, which can facilitate future replications and extensions of this work.}
\end{itemize}

\section{Background and Terminology}
\label{sec:background} 

\begin{figure}[t]
\centering
\footnotesize
\makebox[\columnwidth]{
\begin{minipage}[t]{\columnwidth}
\begin{bugreport}{Bug Report}{}
\textbf{Title:}\\
Bug: Long pressing the amount input brings up QWERTY keyboard\\
\textbf{Content:}\\
\textit{Software specifications:}\\
$\sbullet$ GnuCash Android version: 2.2.0\\
$\sbullet$ System Android version: 6.0\\
$\sbullet$ Device type: Motorola Moto G (2nd Generation)\\
\textit{Steps to reproduce the behaviour:}\\
1. Navigate to Transactions screen\\
2. Tap the Add button\\
3. Enter Description (optional)\\
4. Focus the Amount input\\
5. Long press to bring up the context menu\\
\textit{Expected behaviour:}\\
See the context menu\\
\textit{Actual behaviour:}\\
\begin{center}
\vspace{-12pt}
\adjincludegraphics[width=0.6\textwidth,valign=t,trim={0 {.65\height} 0 {.037\height}},clip]{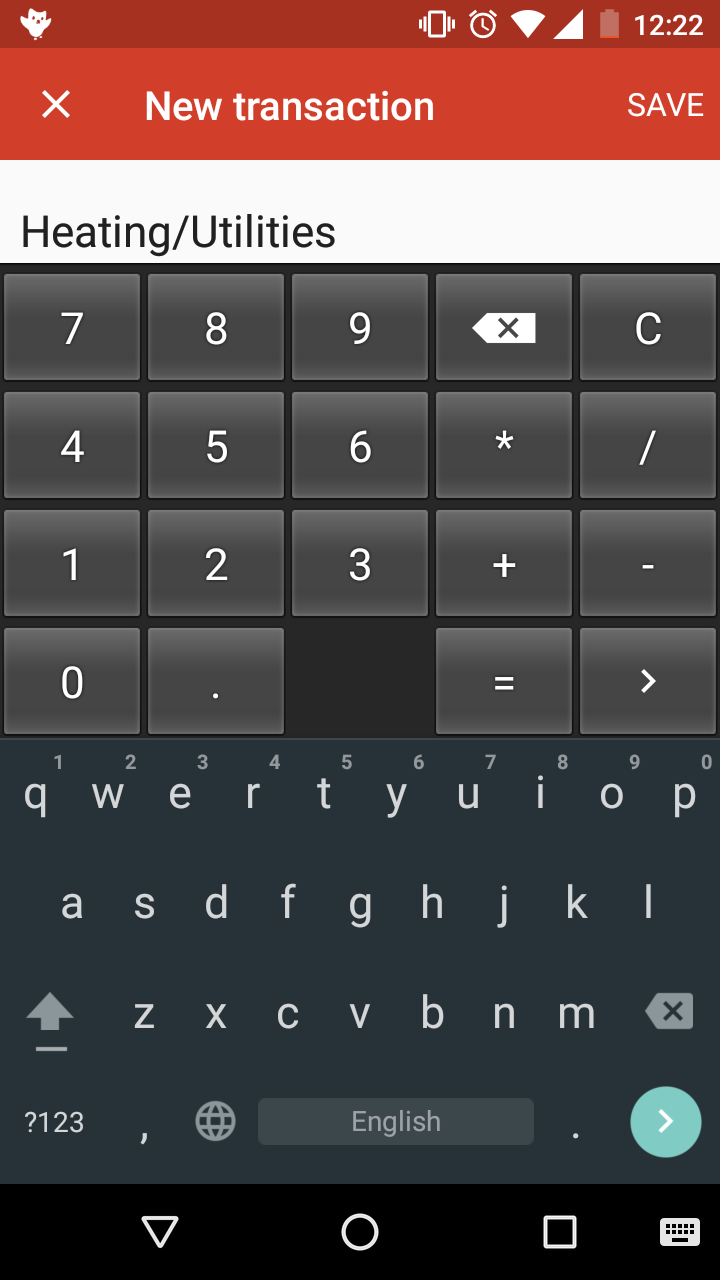}
\end{center}
\end{bugreport}
\vspace{5pt}
\caption{Bug report for the \textsc{GnuCash} app.}
\label{fig:bugreport}
\vspace{-22pt}
\end{minipage}
}

\end{figure}

Given a bug report that describes a failure in an app, we use the term \textit{reporter} to identify the person submitting the bug report. A reporter can be either a \textit{user} or a \textit{developer}. 
In this study, we consider a person who never contributed to the source code of an app to be a user and 
all other reporters to be developers.  

We conceptually group the information contained in a bug report into multiple parts, each of which detail a particular aspect of the report. The parts and aspects of interest in this study are the ones providing details on how to reproduce the failure described in a report. These aspects are: the \textit{environment}, the \textit{steps to reproduce} (S2Rs), and the \textit{observed behavior} (OB). The environment includes information on the software and hardware necessary to reproduce the failure described in a report. This part can contain information such as the app version, the operating system (OS) version, and the device where the failure occurred. The S2Rs provide details on the operations that should be performed on a device in order to reproduce the failure. We use the terms \textit{GUI action} (or simply \textit{action}) and \textit{GUI interaction} (or simply \textit{interaction}) interchangeably to indicate the operations performed on the GUI of a device. An S2R (which are the unit of information composing the S2Rs) can be mapped to one or more GUI actions. The OB describes the failure and can be used to check that the failure was successfully reproduced. 
In practice, the information from these conceptual parts can be interleaved across the paragraphs and sections of a bug report. Bug reports can also have a \textit{discussion thread}. A discussion thread contains \textit{discussion messages} and these messages can provide additional information on the environment, the S2Rs, and the OB associated with the report.

Figure~\ref{fig:bugreport} provides an example of a user-submitted bug report~\cite{2021_github_gnucash_689}.
This bug report is taken from the report management system of \textsc{GnuCash}, an app for finance tracking, and 
is slightly modified for presentation purposes. 
The bug report contains information related to the environment, the S2Rs, and the OB, which are located in the \textit{Software specifications}, \textit{Steps to reproduce the behaviour}, and \textit{Actual behaviour} sections of the report, respectively. 

To exercise the bug, the user navigated to the transactions screen, started adding a new transaction, and long-clicked on the GUI element representing the amount of the transaction. The failure manifests as a wrong screen being displayed to the user: screen with a keyboard view instead of the context menu.
The OB describing the failure is reported using text (in the title) and using an image (in the \textit{Actual behaviour} section). We refer to the way in which a piece of information is reported as the \textit{reporting modality} (or \textit{modality} in short) and reporters can provide the same information multiple times using different modalities.
Because the user did not reach the desired screen, we identify this failure as a \textit{navigation failure}. We use the terms \textit{failure type} and \textit{failure category} interchangeably to refer to the categorization of the failure.

The report has five S2Rs (numbered items under the \textit{Steps to reproduce the behaviour} section) and 13 GUI actions are necessary to reproduce the failure. An example of GUI action is performing a click on the add button in the transaction screen of the app as indicated by \textit{2. Tap the Add button}. 
An S2R can map to one or more GUI actions. In this example, the first S2R (\textit{1. Navigate to Transactions screen}) maps to three GUI actions. We refer to S2Rs that map to multiple GUI actions as \textit{non-specific S2Rs}. Of the remaining
four S2Rs, three map to one GUI action and one S2R is optional (\textit{3.  Enter Description (optional)}.) This optional S2R is not included in 13 GUI actions necessary to reproduce the failure. Seven (13-3-3) of the GUI actions in this example are not described
by any of the S2Rs.
We refer to such GUI actions as \textit{unmapped GUI actions} and say that they correspond to \textit{missing S2Rs}.
We refer to the remaining actions as \textit{mapped GUI actions}. 
If an unmapped GUI action occurs before the first mapped GUI action, we call the missing S2R that corresponds 
to the unmapped action a \textit{missing context S2R}, indicating that some contextual information is missing from the bug report. Otherwise, if a missing S2R is associated with a GUI action occurring after the first mapped GUI action, we refer to the S2R as a \textit{missing inline S2R}.
\begin{figure*}[t!]
	\centering
	\vspace{-1em}
	\centerline{\includegraphics[width=1.0\textwidth]{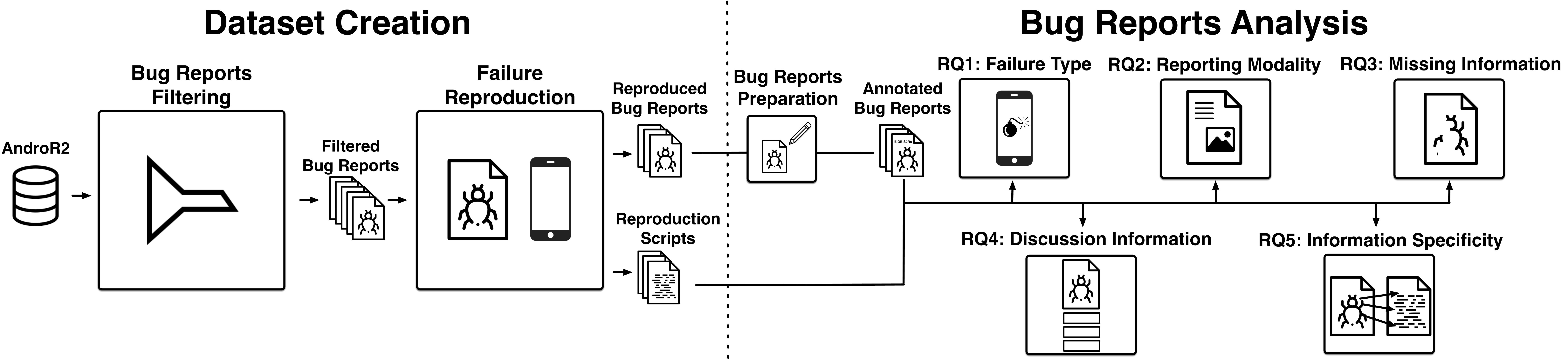}}
	\caption{Overview on the methodology used in the study.} \label{fig:methodology}
	\vspace{-16pt}
\end{figure*}

\section{Methodology}

To characterize reproducible bug reports, inform research on automated bug reproduction, and, more generally, provide insights for research on bug report management, we formulated and answered the following research questions (RQs):

\begin{itemize}
\setlength{\itemindent}{-5pt}
\item \textbf{RQ$_1$: What are the failure types associated with reproducible bug reports?}
In this RQ, we analyzed and categorized failures associated with reproducible bug reports.
With the findings from this RQ we aim to inform research on automatic failure recognition.

\item \textbf{RQ$_2$: What information modalities are used to report the information contained in reproducible bug reports?}
This RQ categorizes the modalities used to report environment, S2Rs, and
OB information.
The findings from this RQ aim to inform research in bug triaging, report reproduction, and report quality assessment.

\item \textbf{RQ$_3$: Do reproducible bug reports have missing information?}
We answer
this question by analyzing the information contained in reproducible bug reports w.r.t. operations required to reproduce the failures described in the reports.
This RQ aims to direct efforts on research for identifying and inferring missing information
in bug reports, necessary for bug report reproduction.

\item \textbf{RQ$_4$: Do discussion threads of reproducible bug reports contain helpful information for reproducing the reports?}
In this RQ, we analyzed the information gain obtained by interpreting the bug report discussions.
This RQ aims to evaluate the need for approaches that combine content from bug reports and their discussions.

\item \textbf{RQ$_5$: How specific is the information reported in reproducible bug reports?} 
In this RQ, we investigated whether the information contained in reproducible bug reports can be directly mapped
onto the operations need to reproduce the reports. This RQ aims to provide insights on how to
leverage the information in bug reports for reproducing the failures.
\end{itemize}

Figure~\ref{fig:methodology} provides a high-level outline of the methodology we used to answer the RQs. 
In a nutshell, we first assembled a dataset of reproducible bug reports and then analyzed the characteristics of the bug reports through qualitative and quantitative analyses.
We describe these steps in detail next. 

\subsection{Dataset Creation}

The \textit{Dataset Creation} component of Figure~\ref{fig:methodology} provides an overview of our data 
collection workflow, which consisted of two phases: 
\textit{bug reports filtering} and \textit{failure reproduction}.

\subsubsection{Bug Reports Filtering}

The objective of this phase was to identify a set of bug reports that we could try to reproduce and ultimately 
include in our dataset. In this study, we are interested in both user-submitted and developer-submitted bug reports that are reproducible and describe different types of failures. To the best of our knowledge, \textsc{AndroR2}~\cite{Wendland2021} is the largest dataset of reproducible bug reports for Android apps that does not exclusively focus on crashes. This dataset contains 90 user-submitted bug reports, which are associated with apps available on the Google Play store~\cite{2021_google_play} and hosted on GitHub~\cite{2021_github}. 
The 90 bug reports are GitHub issues~\cite{2021_github_issues} and are associated with reproduction scripts created by the \textsc{AndroR2}'s authors. This set of 90 bug reports was extracted from a larger set of 6,365 issues that was systematically mined from GitHub. The set of 6,365 issues contains issues that: (i) are part of repositories that use Java, (ii) have the label ``bug'', (iii) are in repositories that contain an \texttt{\small AndroidManifest.xml} file (as Android apps require this file to properly compile~\cite{2021_app_manifest_overview}), (iv) contain the word ``step'' in them, and (v) are associated with apps also available on the Google Play store.

Because we are also interested in developer-submitted bug reports, we started from the set of 6,365 GitHub issues provided by \textsc{AndroR2} and identified 90 reproducible, developer-submitted bug reports (to match the number of already available user-submitted bug reports). To identify the 90 developer-submitted bug reports, we used a methodology similar to that of \textsc{AndroR2}.
Specifically, we first refined the set of 6,365 issues to only contain those created by GitHub users that had contributed to the repositories associated with the issues, resulting in 2,523 issues. 
Second, we selected issues that were closed at the time the issues were mined (November 2020) so that we could more easily identify whether the issues were also originally reproduced by the developers.
This filtering resulted in 2,045 reports.
Third, after analyzing the set of issues, we found that some repositories had a much larger number of issues compared to others. To avoid overfitting the bug report dataset to a specific app, we considered at most ten issues per repository. When a repository had more than ten issues, we randomly selected ten
from this set resulting in 645 bug reports for 164 apps.

\subsubsection{Failure Reproduction Phase}

In the second phase of our dataset creation process, we randomly selected bug reports from the set of 645 developer-submitted bug reports until we reproduced 90 of them. In this process, we disregarded trivially reproducible bug reports, i.e., those we could reproduce by simply opening the app.

Two authors tried to reproduce the failures described in the bug reports. To reproduce a failure, the authors followed the S2Rs contained in the bug report by mapping the steps to GUI actions on the screen of the device running the app associated with the report. If a report had missing S2Rs, the authors manually explored the functionality of the app to identify the minimal sequence of GUI actions that would account for those missing steps, using a trial-and-error approach. When a bug report could be successfully reproduced by one of the two authors, the other author also tried to reproduced the same report to ensure that the reproduced failure was the same as the one described in the report.  For all 90 bug reports, the authors also encoded the GUI actions in reproduction scripts using the UIAutomator framework~\cite{2021_google_uiautomator}.

To validate whether user-submitted bug reports were still reproducible, we ran the scripts associated with these reports in the \textsc{AndroR2} dataset. Four reports were not reproducible as the servers associated with the apps were no longer running. To replace these bug reports, we identified and reproduced four additional user-submitted reports from the set of 6,365 GitHub issues provided by \textsc{AndroR2}. 
At the end of this process, we obtained a set of 90 user-submitted and 90 developer-submitted reproducible bug reports, which we considered for the rest of the study.

\subsection{Bug Reports Analysis}

In this section, we present the analyses we performed to characterize aspects related to the reproducibility of Android bug reports. The \textit{Bug Reports Analysis Creation} part of Figure~\ref{fig:methodology} provides a summary of the analyses we performed. The analyses were driven by two of the paper's authors and were performed one at a time to reduce cognitive load.

\subsubsection{Bug Reports Preparation}

Before performing the analyses associated with the RQs, we annotated the information contained in the bug reports and their discussion threads, to identify the portions of each report that provide information about the environment, S2Rs, and OB. 
This step was performed by the two authors together and in multiple sessions;
the authors associated each sentence in the report's textual description, as well as each link, image, recording, and execution logs, with it designated purpose: to describe environment, S2Rs, and OB. 
Some elements received multiple annotations, \eg a sentence can provide both S2Rs and OB.

\subsubsection{Analysis for RQ$_1$ (What are the failure types associated with reproducible bug reports?)}
To answer RQ$_1$, we performed a qualitative analysis that combines inductive and axial coding ~\cite{2014_sage_corbin_basics, 2018_sage_miles_qualitative}. Inductive coding is a systematic approach for categorizing data by manually coding (\ie labeling) the data. Axial coding relates codes to one another and finds higher-level codes that represent abstractions of the original codes. In our analysis, a code is a label that categorizes the type of a failure and we assigned the code to the bug report describing the failure.

The analysis was performed by two raters, who analyzed the description of the failure in the bug report and used the reproduction scripts to observe how the failure manifested. The analysis was divided into two parts. In the first part, the two raters analyzed a sample of the bug reports to define the analysis codebook -- a document detailing the rules for assigning a specific code to a failure. For each code, the set of rules specified the characteristics required for assigning a code to a failure.  

This part of the analysis was performed in six iterations. In each iteration, the raters independently analyzed 18 bug reports (10\% of the report considered in the study). The set contained the same bug reports for both raters and was selected randomly from the set of not-yet-analyzed bug reports. At the end of each iteration, the raters used negotiated agreement~\cite{2013_campbell_coding} to resolve inconsistencies among created and assigned codes, and to insure the reliability of the coding process. We used this method due to it is advantages in research like ours, where generating new insights is the primary concern~\cite{1974_morrissey_sources}. Because we used negotiated agreement, measures such as inter-rater agreement are not applicable in our context. To resolve disagreements, the raters reproduced the failures together and then decided on the final classification. For example, for one of the reports considered in the study~\cite{calendula}, one of the raters categorized the failure as a crash and the other rater categorized the failure as a navigation issue. When the two raters met, they discussed the disagreement and decided to classify the failure as a crash because the app displayed an exception before bringing the user back to a different screen.

At the sixth iteration, the raters did not create new codes and had assigned the same codes to all reports. From that point, the raters split the remaining 72 bug reports equally and coded the bug reports independently.
At the end of the coding process, the raters also performed axial coding. This step led to four main categories of failures, which we present in Section~\ref{sec:results}.

\subsubsection{Analysis for RQ$_2$ (What information modalities are used to report the details contained in reproducible bug reports?)}

The analysis to answer RQ$_2$ was also based on inductive and axial coding. 
Two raters analyzed the environment, S2Rs, and OB information annotated during the bug reports preparation step. The raters created the analysis codebook in two iterations, analyzing in each iteration a sample of 18 bug reports (10\% of all bug reports). The raters used negotiated agreement to address the reliability of the coding process.  After finalizing the codebook, the authors split the remaining 144 bug reports equally and coded them independently.

The raters performed axial coding at the end of the coding process. This process led to six main reporting modalities, detailed in Section~\ref{sec:results}.

\subsubsection{Analysis for RQ$_3$ (Do reproducible bug reports have missing information?)}

To answer RQ$_3$, we performed two types of analysis. First, we leveraged the annotations created in the bug reports preparation step to identify whether environment, S2Rs, and OB information was completely missing from the reports. Second, when the S2Rs information was provided, we performed an in-depth analysis of S2Rs. Specifically, for each bug report, we compared the S2Rs information from the bug report with the GUI actions in our reproduction scripts, in order to identify missing S2Rs. Once we identified missing S2Rs, we categorized them into missing context S2Rs and missing inline S2Rs (see definitions in Section~\ref{sec:background}). Two authors analyzed each bug report independently and then met to discuss and finalize the classification.

\subsubsection{Analysis for RQ$_4$ (Do discussion threads of reproducible bug reports contain helpful information for reproducing the reports?)}

In RQ$_4$, two authors manually analyzed the messages in the bug report discussions, to identify whether they added information relevant to understanding and reproducing the bug reports. The authors leveraged the annotations from the bug reports preparation step to focus on messages providing environment, S2Rs, and OB information. The authors analyzed each bug report independently and labeled with the word \textit{additional} the data from discussion messages that provided additional information. The two authors met and discussed the final classification also in this case.

\subsubsection{Analysis for RQ$_5$ (How specific is the information reported in reproducible bug reports?)}
 
To answer RQ$_5$, we analyzed whether the information provided in the bug reports could be directly used for reproducing the bug reports. For the environment-related information, two authors checked whether the provided information was sufficient to define the environment where to reproduce the failure. If no additional information was needed, we considered the provided information to be of specific (and non-specific otherwise). For S2Rs, two authors mapped each of the S2Rs defined in a bug report to corresponding GUI actions from the reproduction script. If an S2R mapped to multiple GUI actions, we labeled that S2R as a non-specific S2R. We considered the other S2Rs to be specific.
For the OB information, the authors checked whether the information was sufficient to verify the failure. If no additional information was needed (\ie no need to check discussion messages), we considered the provided information to be specific (and non-specific otherwise). 
\section{Results}
\label{sec:results}

In this section, we present the results of our study on analyzing
and characterizing reproducible Android bug reports.

\begin{figure*}
\vspace{-1em}
    \centering
    \begin{subfigure}[b]{0.42\textwidth}
	\adjincludegraphics[width=0.49\textwidth,valign=t,trim={0 {.6\height} 0 {.035\height}},clip]{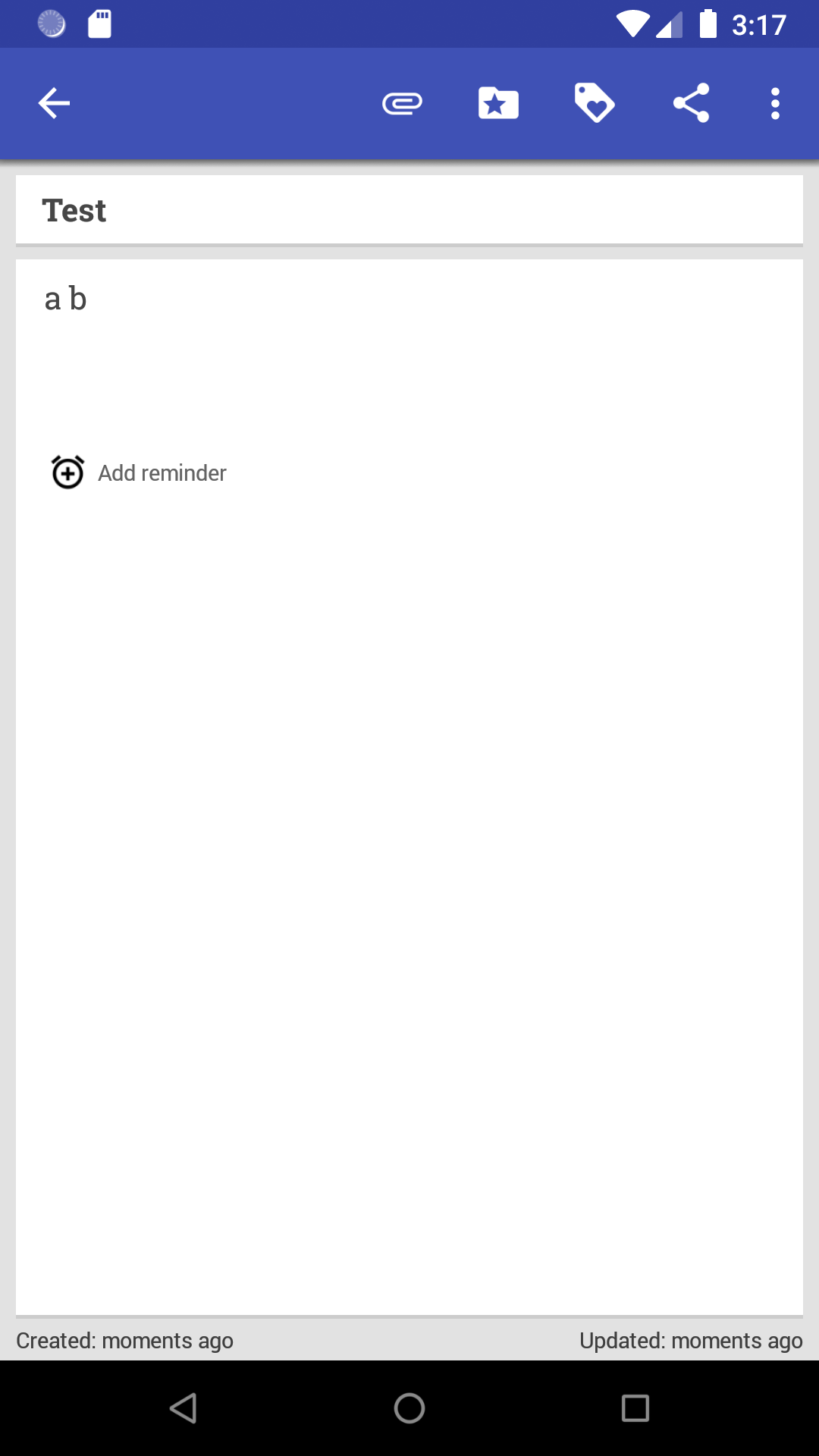}
	\adjincludegraphics[width=0.49\textwidth,valign=t,trim={0 {.6\height} 0 {.035\height}},clip]{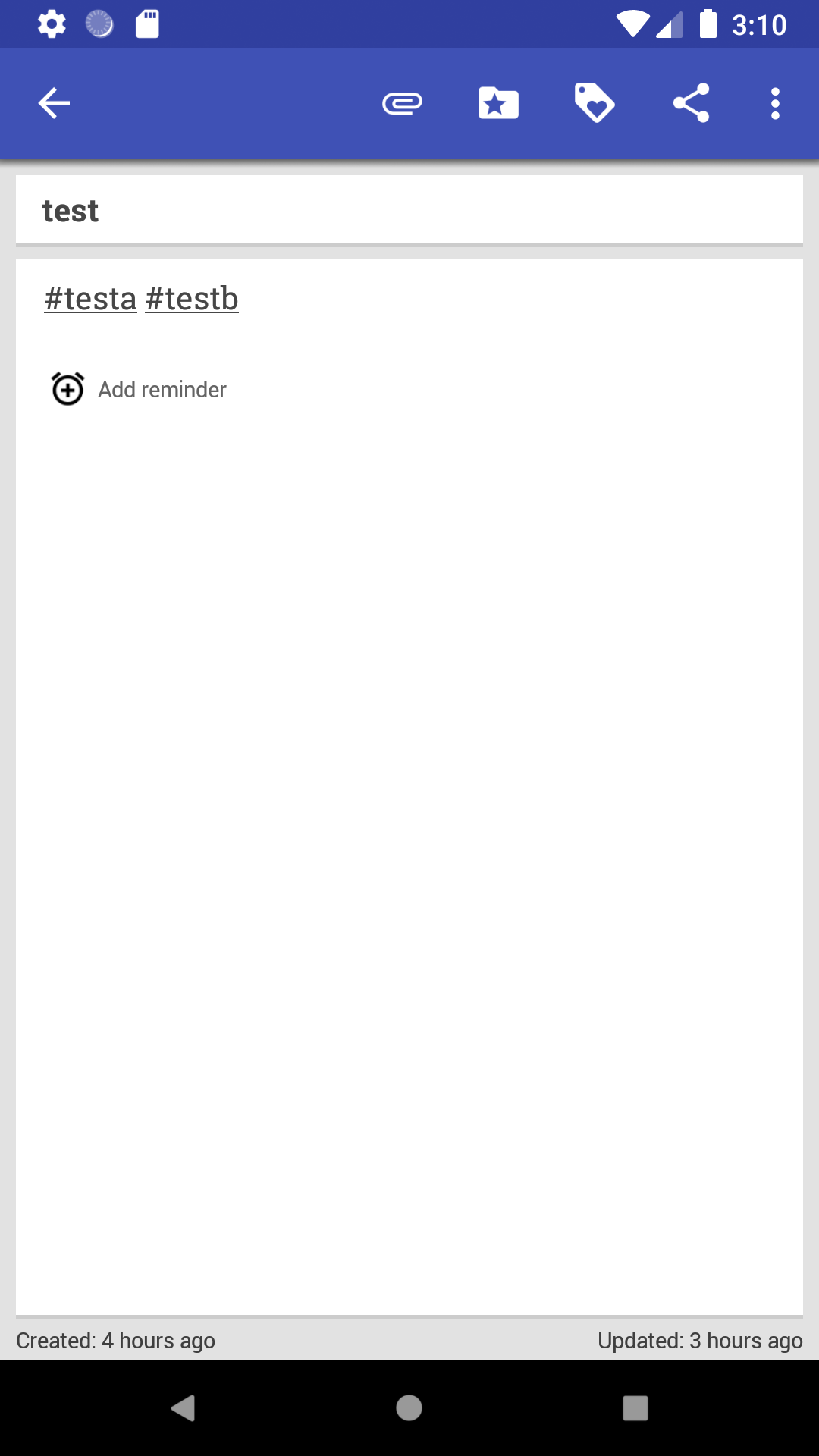}
	\adjustbox{center,minipage=1.1\textwidth}{\vspace{5pt}\caption{Example of output failure on the left and fix on the right.\label{fig:output}}}
    \end{subfigure}
    \hspace{20pt}
    \begin{subfigure}[b]{0.42\textwidth}
	\adjincludegraphics[width=0.49\textwidth,valign=t,trim={0 {.6\height} 0 {.035\height}},clip]{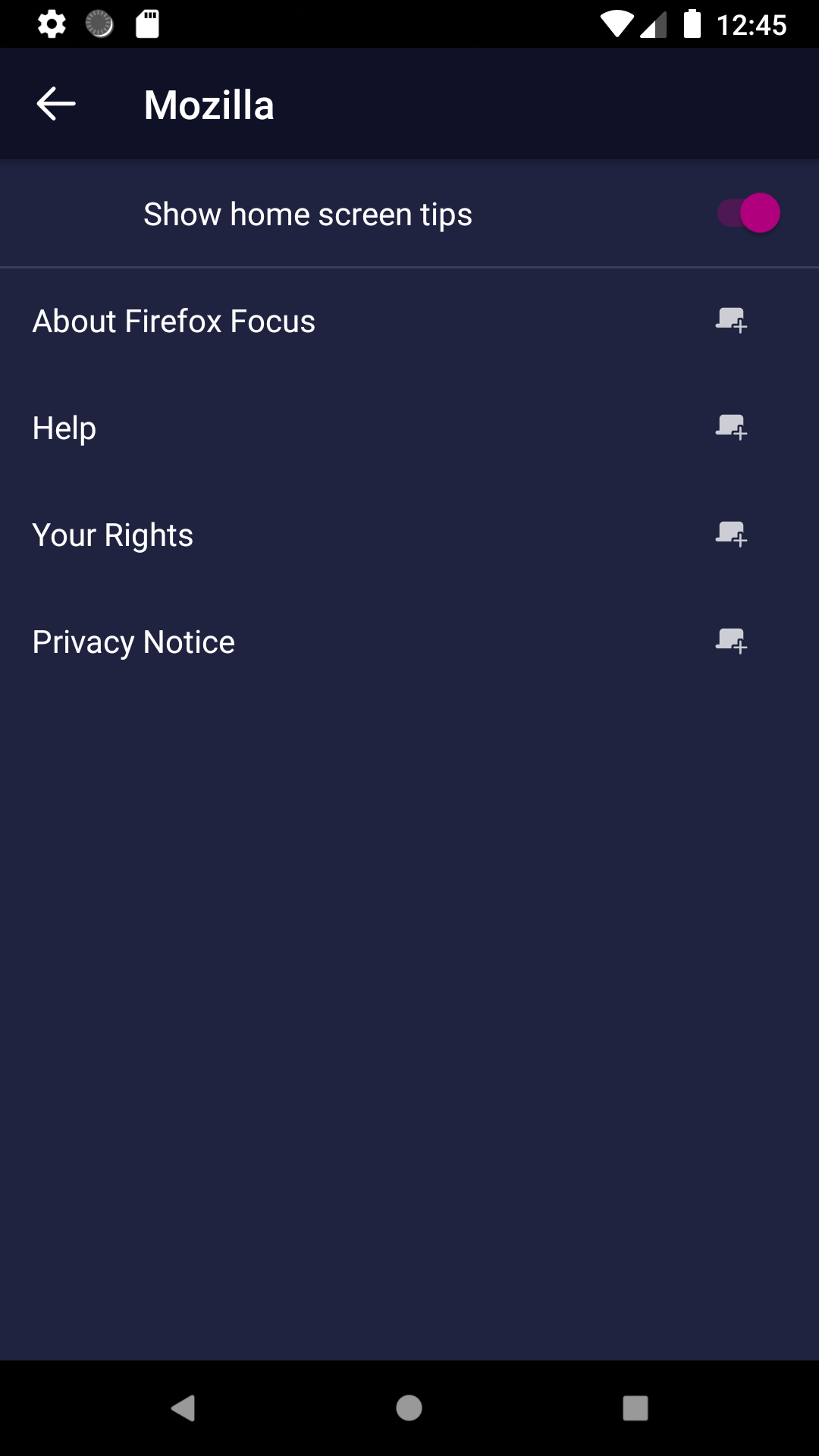}
	\adjincludegraphics[width=0.49\textwidth,valign=t,trim={0 {.6\height} 0 {.035\height}},clip]{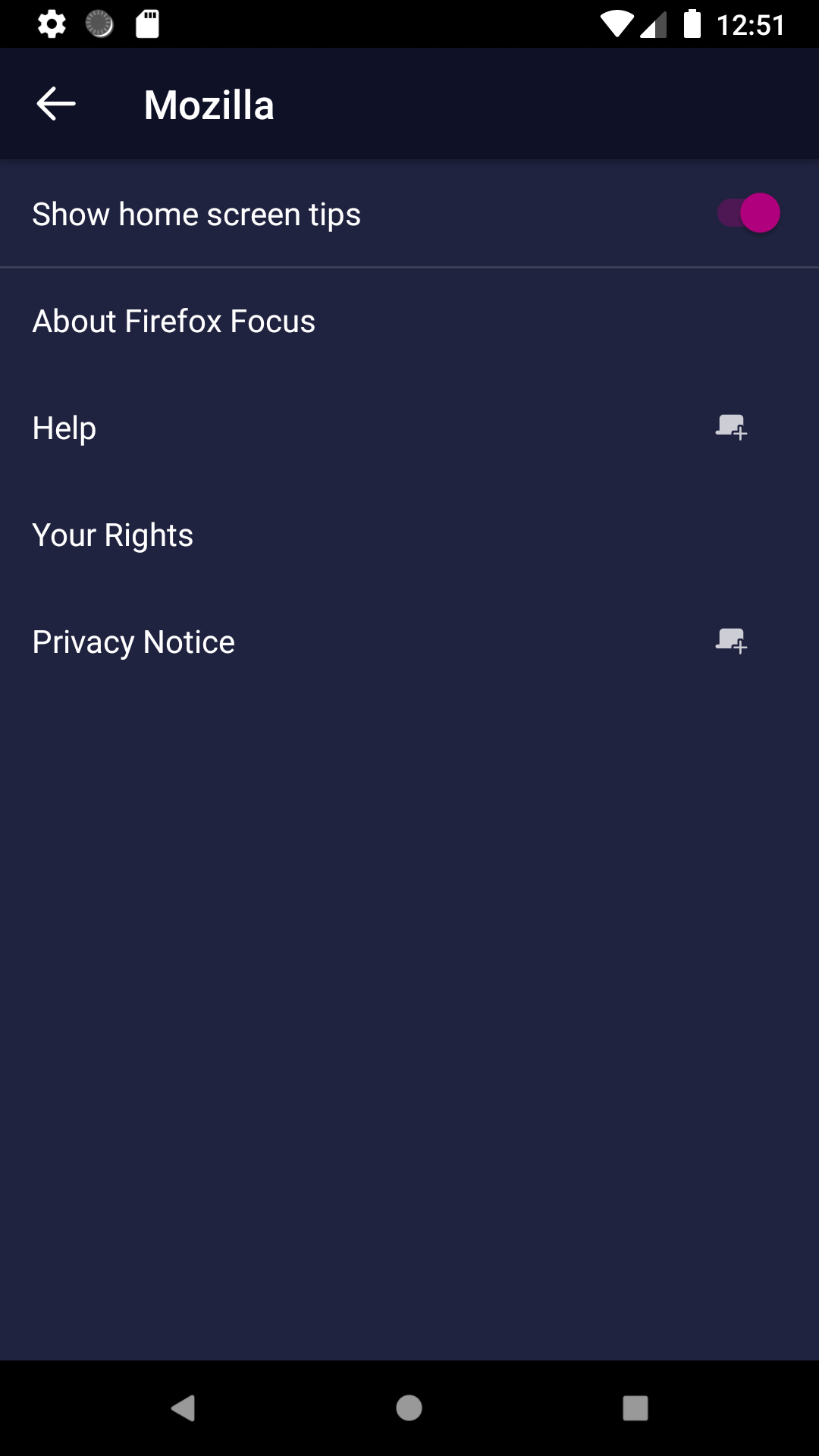}
     \adjustbox{center,minipage=1.1\textwidth}{\vspace{5pt}\caption{Example of cosmetic failure on the left and fix on the right.\label{fig:cosmetic}}}
    \end{subfigure}
    \vspace{20pt}
    \begin{subfigure}[b]{0.42\textwidth}
	\adjincludegraphics[width=0.49\textwidth,valign=t,trim={0 {.6\height} 0 {.035\height}},clip]{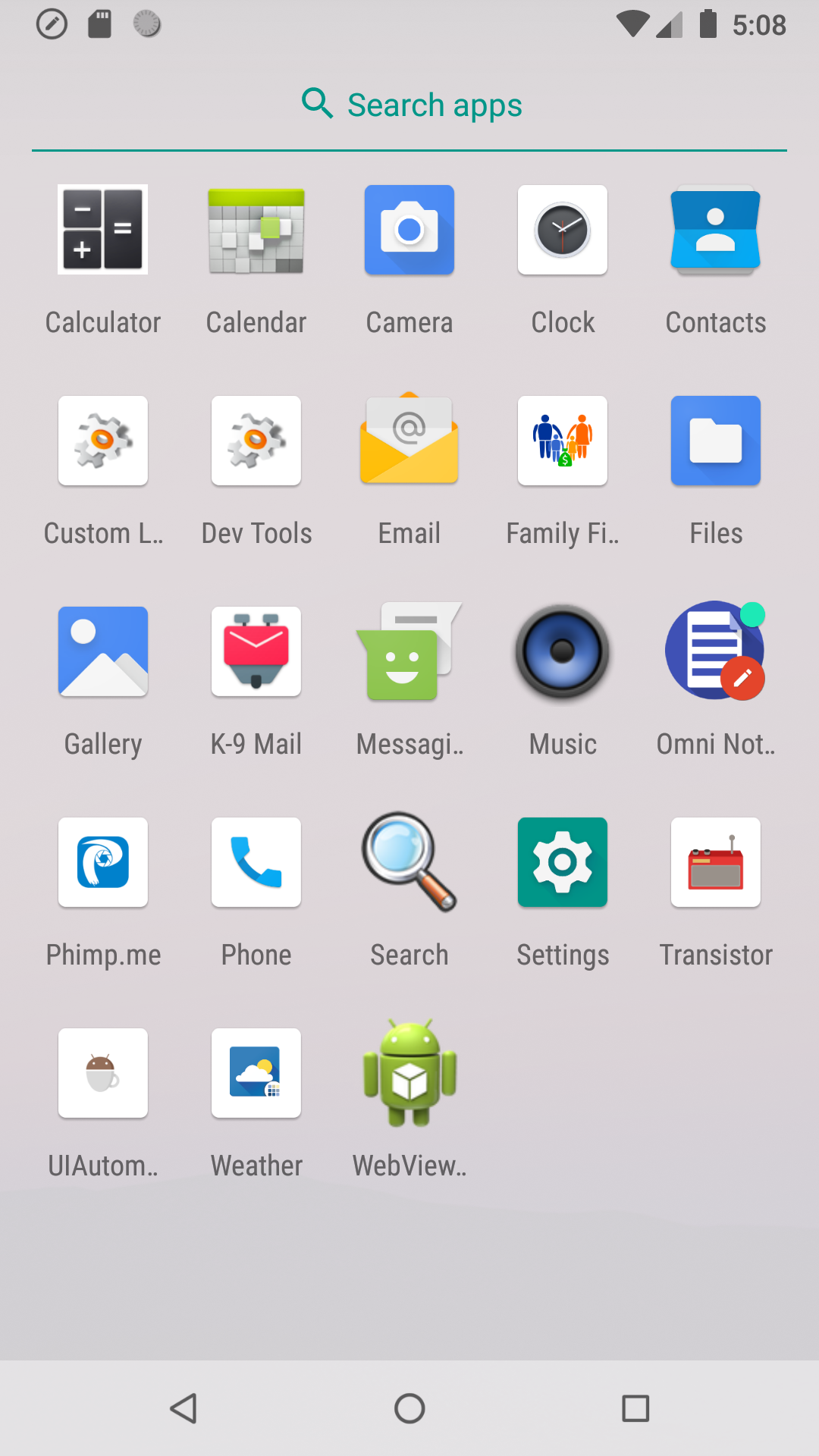}
	\adjincludegraphics[width=0.49\textwidth,valign=t,trim={0 {.6\height} 0 {.035\height}},clip]{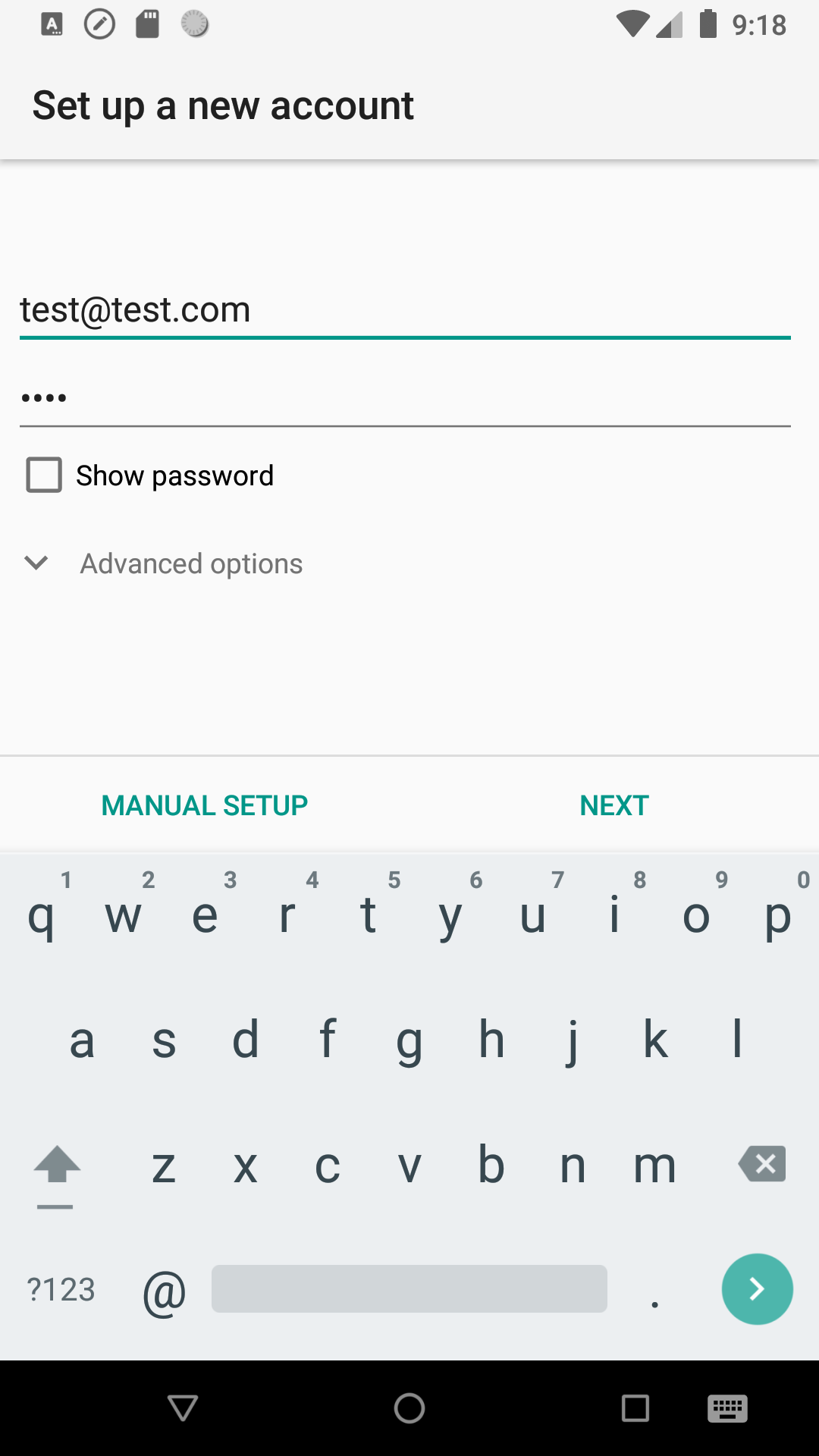}
     \adjustbox{center,minipage=1.1\textwidth}{\vspace{5pt}\caption{Example of navigation failure on the left and fix on the right.\label{fig:navigation}}}
    \end{subfigure}
    \hspace{20pt}
    \begin{subfigure}[b]{0.42\textwidth}
	\adjincludegraphics[width=0.49\textwidth,valign=t,trim={0 {.5\height} 0 {.135\height}},clip]{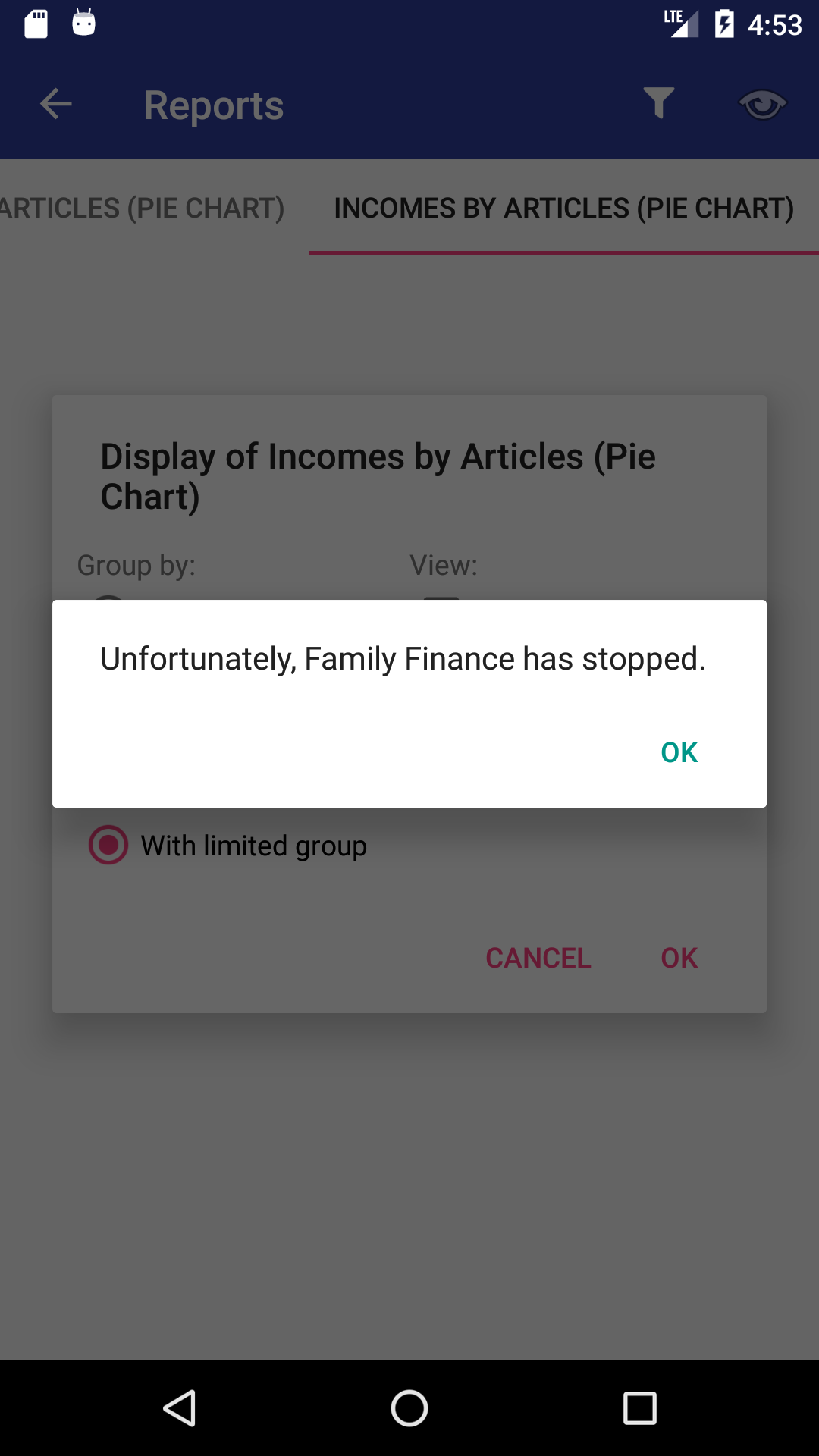}
	\adjincludegraphics[width=0.49\textwidth,valign=t,trim={0 {.5\height} 0 {.135\height}},clip]{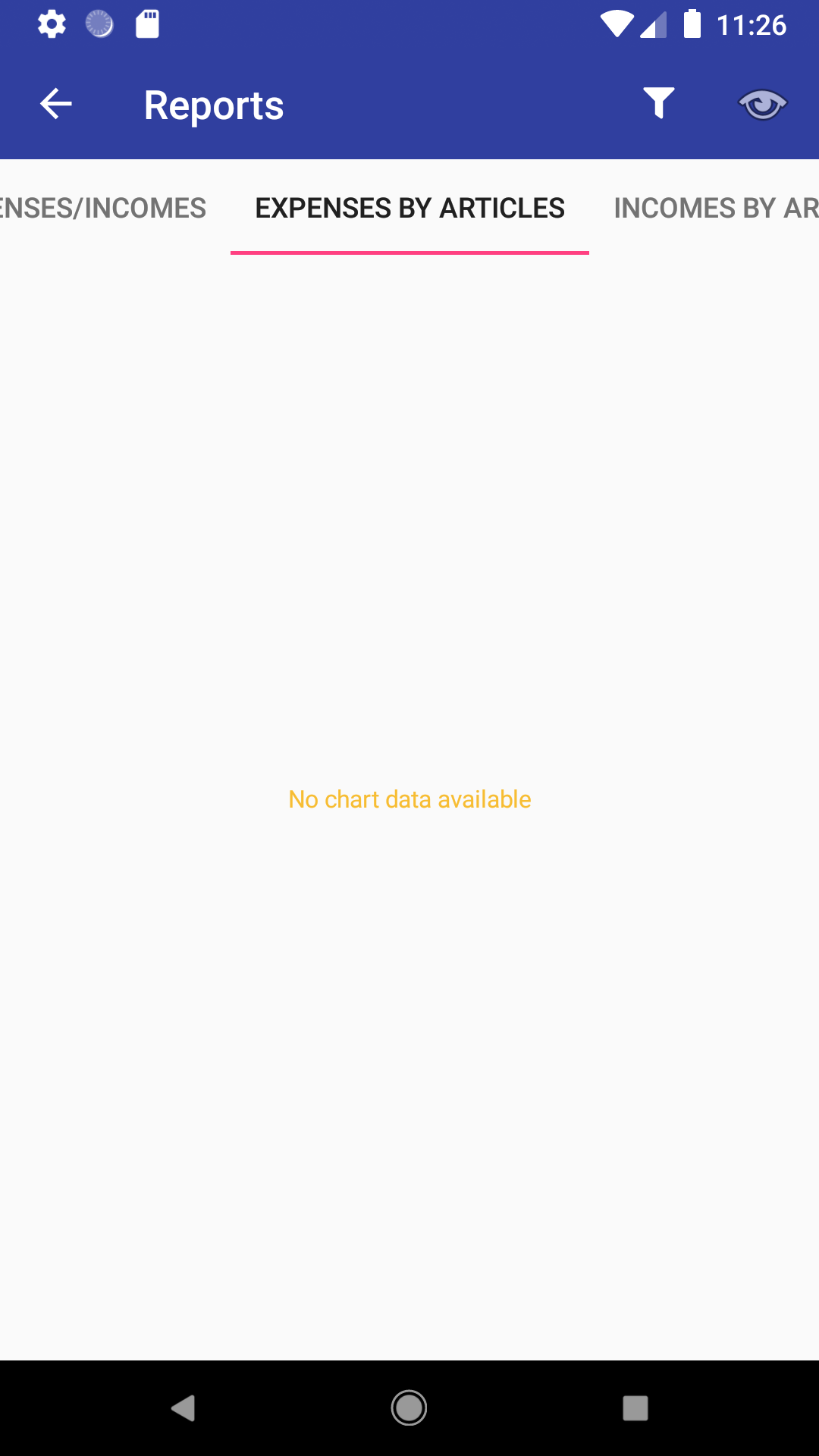}
     \adjustbox{center,minipage=1.1\textwidth}{\vspace{5pt}\caption{Example of crash failure on the left and fix on the right.\label{fig:crash}}}
    \end{subfigure}   
    \vspace{-25pt}
    \caption{Screenshot examples for the four failure types identified in the bug reports considered.}
    \label{fig:RQ1_examples}
    \vspace{-15pt}
\end{figure*}

\subsection{RQ$_1$: What are the failure types associated with reproducible bug reports?}

Our analysis identified four failures types: \textit{output}, \textit{cosmetic}, \textit{navigation}, and \textit{crash}. Output failures reveal issues in the output provided by the app. Cosmetic failures identify issues in the app that do not affect the functionality of the app. Navigation failures display the wrong screen to the user. Crashes abruptly terminate the execution of the app. Across the bug reports considered, we identify 33\% of reports reporting output failures, 31\% reporting cosmetic failures, 8\% reporting navigation failures, and 28\% reporting crashes. This finding is notable, as many current bug report analysis techniques focus solely on crashes. We discuss the implications of these findings further in Section~\ref{sec:discussion}.

This distribution reveals a comparable amount of failures between the output, cosmetic, and crash categories and a significantly lower number of navigation failures. The distribution is similar across both developer- and user-submitted bug reports. Specifically, among the user-submitted bug reports, there are 33\% output failures, 31\% cosmetic failures, 7\% navigation failures, and 28\% crashes. Among developer-submitted bug reports, there are 32\% output failures, 29\% cosmetic failures, 9\% navigation failures, and 29\% crashes. 

Our analysis categorized the 60 output failures into two subcategories: \textit{incorrect output} (32) and \textit{missing output} (28). \textit{Incorrect output} identifies failures in which some computation of the app is displayed incorrectly or improperly saved to a file, and \textit{missing output} describes failures where the result of some computation is not displayed or saved to a file. A vast majority of these cases affect the GUI of the app (56 cases) whereas a smaller number impact generated files (4 cases).

The screenshot on the left of Figure~\ref{fig:output} shows an example of a failure under the incorrect output subcategory. The example is taken from a bug report~\cite{2021_github_omni-notes_634} of \textsc{Omni Notes}, a note-taking app. The app has a failure as it does not display the right values for the tags associated with the notes in the app.

As part of our analysis, we further classified the 55 cosmetic failures into eight subcategories: \textit{incorrect color} (10), \textit{incorrect cursor placement} (3), \textit{content cut} (3), \textit{image rendering issue} (4), \textit{missing GUI element} (9), \textit{incorrect orientation} (2), \textit{incorrect placement} (4), and \textit{incorrect text} (18). We provide details for each of these subcategories in our online appendix~\cite{appendix}.
The screenshot on the left of Figure~\ref{fig:cosmetic} illustrates an example of a cosmetic failure from the \textit{incorrect placement} subcategory. This example is taken from a report~\cite{2021_github_firefox-focus_3304} submitted for \textsc{Firefox Focus}, a browser app. In this example, the text {\small\texttt{Show home screen tips}} has additional padding w.r.t other text elements (\eg {\small\texttt{ About Firefox Focus}}) on the screen.

Our analysis of the navigation failures did not produce any further subcategories. The screenshot on the left of Figure~\ref{fig:navigation} reports an example of a navigation failure. This failure was reported~\cite{2021_github_k-9-mail_3971} for \textsc{K-9 Mail}, an email client app. In this example,  the user started setting up a new email account, went into the manual configuration settings, and, upon pressing the back button, the user was brought out of the app instead of the previous app screen. The screenshot in the right part of Figure~\ref{fig:navigation} illustrates the correct app behavior where the user navigates to the sign-up screen after pressing the back button.

For the 50 failures leading to a crash, we identified two main subcategories, \textit{immediate crash} (46) and \textit{app freeze} (4). Immediate crash identifies failures in which the app crashes as soon an operation is performed in the app. App freeze includes failures in which the app first becomes unresponsive after an operation is performed in the app, and then the crash appears after a certain amount of time. The screenshot in the left portion of Figure~\ref{fig:crash} reports an example of an immediate crash failure reported~\cite{2021_github_family-finance_1} for \textsc{Family Finance}, a household finance app.
The right part of the Figure~\ref{fig:crash} reports the screen of the app after the bug in the app was fixed.

\begin{tcolorbox}[colback=white, boxsep=0pt, left=4pt, right=4pt, before skip=4pt,after skip=0pt]
\textbf{RQ$_1$ answer:} Our categorization identified four failure types: output (33\%), cosmetic (31\%), navigation (8\%), and crash (28\%). We also identified subcategories for output (2), cosmetic (8), and crash (2). Finally, the failure distribution does not differ dramatically when user- and developer-submitted reports are considered individually. 
\end{tcolorbox}

\subsection{RQ$_2$: What information modalities are used to report the details contained in reproducible bug reports?}

\begin{figure*}[tb]
	\centering
	\vspace{-1em}
	\includegraphics[width=\linewidth]{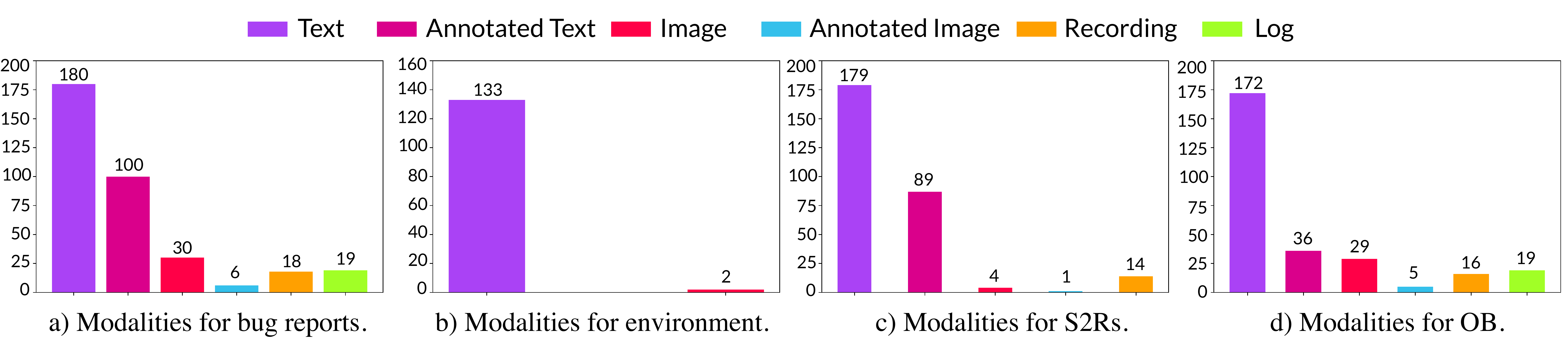}
	\vspace{-0.4cm}
	\caption{Reporting modalities for bug reports and bug report components.}
	\label{fig:rq2}
	\vspace{-0.4cm}
\end{figure*}

In our analysis of RQ2, we identified six modalities used to report bug information: \textit{text}, \textit{annotated text}, \textit{image}, \textit{annotated image}, \textit{recording}, and \textit{log}.
Text identifies information reported in plain text. Annotated text is a  sentence containing text within quotes or text with casing or capitalization~\cite{2021_ux_collective}, which represent either app inputs or GUI elements. Image identifies device screenshots. Annotated image is associated with device screenshots that have been edited to highlight parts of their content. Recording refers to any animated image or video providing a recording of the device screen. Finally, log identifies reporter-provided stack traces extracted from either app or system logs. Figure~\ref{fig:rq2} reports the distribution of the modalities, for reports as a whole (Figure \ref{fig:rq2}-a), the environment (Figure \ref{fig:rq2}-b), S2Rs (Figure \ref{fig:rq2}-c), and OB (Figure \ref{fig:rq2}-d).

As expected, \textit{text} is the most commonly used modality, with all 180 bug reports using text to convey some piece of information. \textit{Annotated text} is the second most recurring modality and appeared in 100 bug reports. In our analysis, we also further categorized the annotated text modality into \textit{annotated GUI text} and \textit{annotated input text}. Annotated GUI text identifies bug reports in which the reporter used text within quotes or latter casing to identify an element in the GUI of the relevant app. An example of this case appears in the bug report associated with Figure~\ref{fig:cosmetic} in which the user wrote \textit{``Show homescreen tips'' is indented} in the report to describe the report's OB. The annotated input text subcategory contains cases in which the reporter provided a textual app input using text within quotes. An example of this case appears in a bug report~\cite{2021_github_k-9-mail_3255} for \textsc{K-9 Mail}, where the reporter mentioned \textit{Add new email account with ``foo@b.2''} as one of the S2Rs in the report. In total, we identified 100 bug reports with annotated text (85 annotated GUI text, six annotated input text, and nine in which both categories appeared).
The remaining modalities, while less common, were still present and, in a large number of cases, provided information that would have been more cumbersome to convey otherwise. Among the bug reports considered, reporters used the image, annotated image, recording, and log modalities in 30, 6, 18, and 19 bug reports, respectively. Furthermore, image-based modalities (\ie image, annotated image, and recording) appeared more frequently in user-submitted (35) than developer-submitted bug reports (14). Finally, we noticed a slight trend of increasing use of image data over the years, with image-based information being present in only ~14\% of reports in 2016 to ~36\% in 2019.

Figures~\ref{fig:rq2}-b, \ref{fig:rq2}-c, and~\ref{fig:rq2}-d report the modalities used for specific sections of the bug reports. Figure~\ref{fig:rq2}-b reports the modalities used for the environment sections. The great majority of the reports (133) use the text modality to report environment information, and only a few use the image modality (2). Figure~\ref{fig:rq2}-c provides the modalities used for the S2Rs. Text is the most commonly used modality (present in 179 bug reports). Annotated text also appears in a considerable number of bug reports (89). The remaining modalities are less common but provide relevant information for reproducing the bug reports. Nineteen bug reports had multiple S2R modalities other than text or annotated text, 16 of these bug reports were user-submitted and 3 were developer-submitted. These 19 bug reports also used the recording (14), the image (4), and annotated image (1) modalities. Finally, Figure~\ref{fig:rq2}-d report the modalities used for the OB sections. Once more, the text modality is the most recurring one (172 cases). However, for OB, the image and recording modalities were used more frequently (29 and 16 cases, respectively) as compared to environment and S2Rs. Sixty bug reports had multiple OB modalities other than text or annotated text, 38 of these bug reports were user-submitted and 22 were developer-submitted. These 60 bug reports also used the image (29), the annotated image (5), recording (16), and log (19) modalities (with some bug reports having multiple modalities). Overall, user-submitted bug reports used reporting modalities other than text more frequently that developer-submitted bug reports.

Examining the relationship between reporting modalities and failure types, we found that bug reports with cosmetic and navigation failures have a higher proportion of cases in which the information is reported using image-based modalities as compared to output and crash failures. Specifically, 45\% of the bug reports describing cosmetic failures and 43\% of the bug reports discussing navigation failures use image-based modalities, while these modalities appear in only 16\% and 18\% of the bug reports describing crash and output failures, respectively. Focusing on specific bug reports sections, we find a similar result for OB descriptions. Additionally, the log modality was used exclusively to report the OB of bug reports describing crashes. These results highlight how certain modalities might be preferable particular failure types.

\begin{tcolorbox}[colback=white, boxsep=0pt, left=4pt, right=4pt, before skip=4pt,after skip=0pt]
\textbf{RQ$_2$ answer:} Our categorization identified six main reporting modalities. Overall, text and annotated text are the most recurring modalities. Certain modalities occur more frequently when considering specific failure types,
e.g., images for cosmetic and navigation failures.
\end{tcolorbox}

\subsection{\small{RQ$_3$: Do reproducible bug reports have missing information?}}

Our analysis identified that 54 bug reports did not contain any environment information, one bug report did not have any S2Rs, and four bug reports did not contain OB information. (Missing information is computed with respect to the bug reports initially submitted and does not consider the information contained in their discussions, as that is the focus of RQ4.)

\begin{figure}[t!]
	\centering
	\vspace{-10pt}
	\includegraphics[width=0.9\columnwidth]{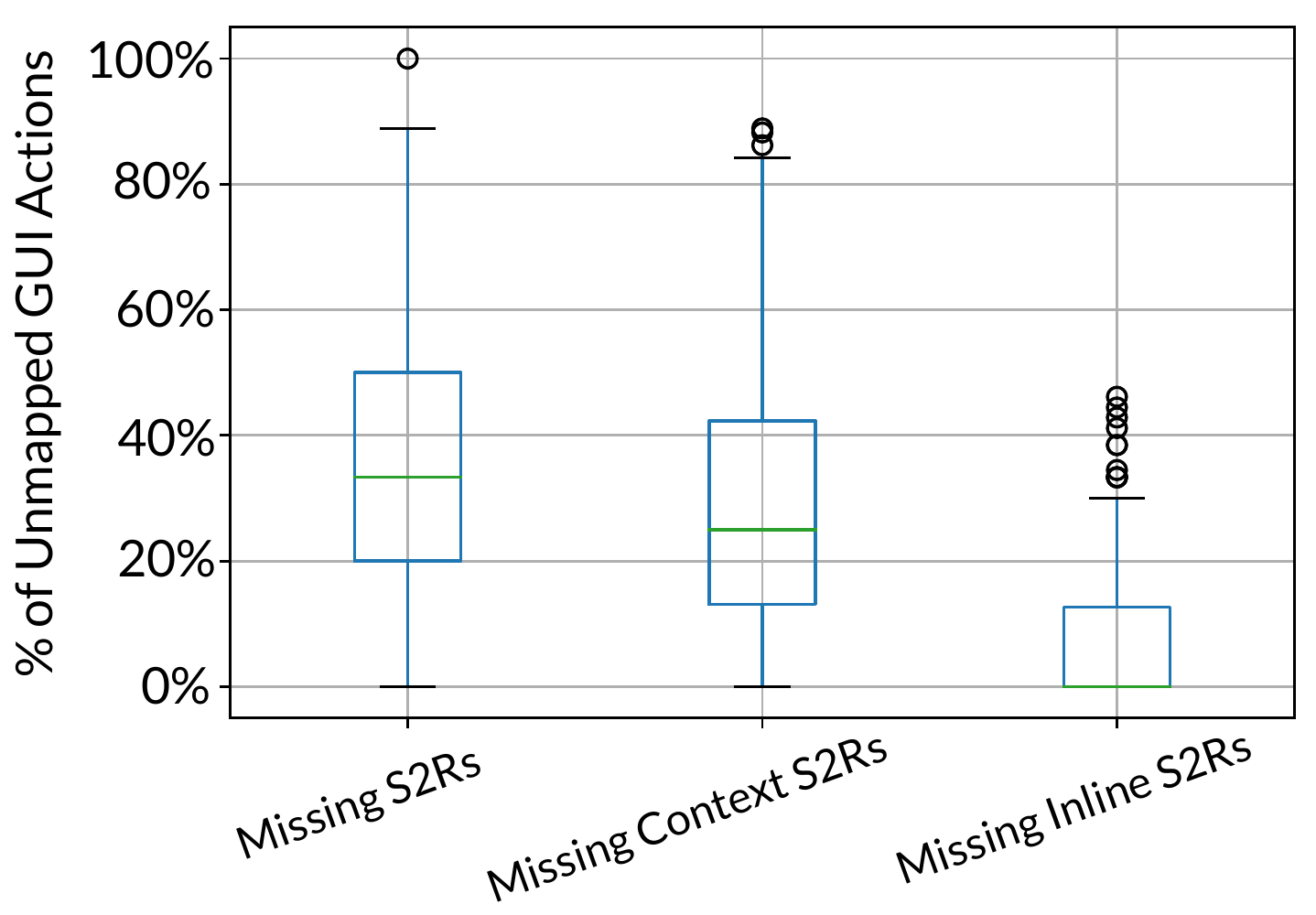}
	\vspace{-5pt}
	\caption{Pct. of unmapped GUI actions due to missing S2Rs.}
	\label{fig:RQ3_results}
	\vspace{-1.5em}
\end{figure}

Although only one bug report did not have any S2Rs, 92.2\% of the bug reports had at least one missing S2R. As mentioned in Section~\ref{sec:background}, missing S2Rs include missing context S2Rs and missing inline S2Rs. 88.3\% of bug reports had at least one missing context S2R and 37.7\% of bug reports had at least one missing inline S2R. Figure~\ref{fig:RQ3_results} associates missing S2Rs to unmapped GUI actions.
More precisely, for each bug report, the figure reports the percentage of unmapped GUI actions with respect to the number of GUI actions necessary to reproduce the report.
The figure reports the percentage for missing S2Rs, missing context S2Rs, and missing inline S2Rs. The figure reveals that 75\% of the bug reports have at least 20\% unmapped GUI actions due to missing S2Rs. Across all bug reports, missing S2Rs led to 43.2\% of GUI actions being unmapped. 33.4\% of unmapped GUI actions are due to missing context S2Rs and 9.8\% are due to missing inline S2Rs. 
These results illustrate that reproducing bug reports also requires inferring a large number of GUI actions that are not specified in the description of the bug reports.

Comparing missing S2Rs from user-submitted bug reports with respect to missing S2Rs from developer-submitted bug reports, users submitted reports that have a lower percentage of unmatched GUI actions due to missing context S2Rs (22.5\%) with respect to developer-submitted reports (43.5\%). This difference does not appear for unmatched GUI actions due to missing inline S2Rs (9.5\% for user-submitted and 10\% for developer-submitted bug reports). We did not observe a difference in missing information across failure types.

\begin{tcolorbox}[colback=white, boxsep=0pt, left=4pt, right=4pt, before skip=4pt,after skip=0pt]
\textbf{RQ$_3$ answer:} The environment section of a bug report is the most likely to be missing from submitted bug reports among the sections considered. A large percentage of bug reports (92\%) had at least one missing S2R. Missing S2Rs equate to 43.2\% unmapped GUI actions necessary to reproduce the failures described in the reports.
\end{tcolorbox}

\subsection{RQ$_4$: Do discussion threads of reproducible bug reports contain helpful information for reproducing the reports?}

To answer this RQ, we analyzed the discussions associated with the bug reports in our dataset and identified information added as part of the conversations that was relevant for reproducing the bugs. In total, 35 of the bug reports contained additional information detailing either the environment, the S2Rs, or the OB of the bug reports. Among these 35 bug reports, 25 were user-submitted and 10 were developer-submitted. Additionally, in 22 of the 35 bug reports, a developer explicitly requested for the information to be added to the discussion.

In the discussions, there were 20 instances of environment information added to the report, 11 instances of S2Rs, and 9 instances of OB. The sum of these numbers is higher than the total number of bug reports with additional information because some discussions (five in total, four with two messages and one with three) contained multiple messages that provided additional information. Although added information does not appear in a large number of cases, these results show that follow up conversations can be leveraged to reproduce reported bugs. Furthermore, considering the high number of reports with missing environment information and unmatched GUI actions identified in RQ$_3$, automated techniques can try to identify and automatically seek this information through iterative or interactive bug reproduction approaches.

Looking at different failure types, bug reports describing output failures were the ones with the highest number of added information in their discussions. Among the 35 bug reports with added information, 17 described output failures, 15 reported crashes, 2 described cosmetic failures, and 1 discussed a navigation failure.

\begin{tcolorbox}[colback=white, boxsep=0pt, left=4pt, right=4pt, before skip=4pt,after skip=0pt]
\textbf{RQ$_4$ answer:} Among the bug reports considered, 35 had additional information relevant for reproducing the reports derived from follow-up, message-based discussions. In 22 reports, the information was explicitly requested by a developer. Finally, of the 35 reports, 20 had added environment info, 11 had added S2Rs, and 10 had added OB.
\end{tcolorbox}

\subsection{RQ$_5$: How specific is the information reported in reproducible bug reports?}

\begin{figure}[t!]
	\centering
	\includegraphics[width=0.9\columnwidth]{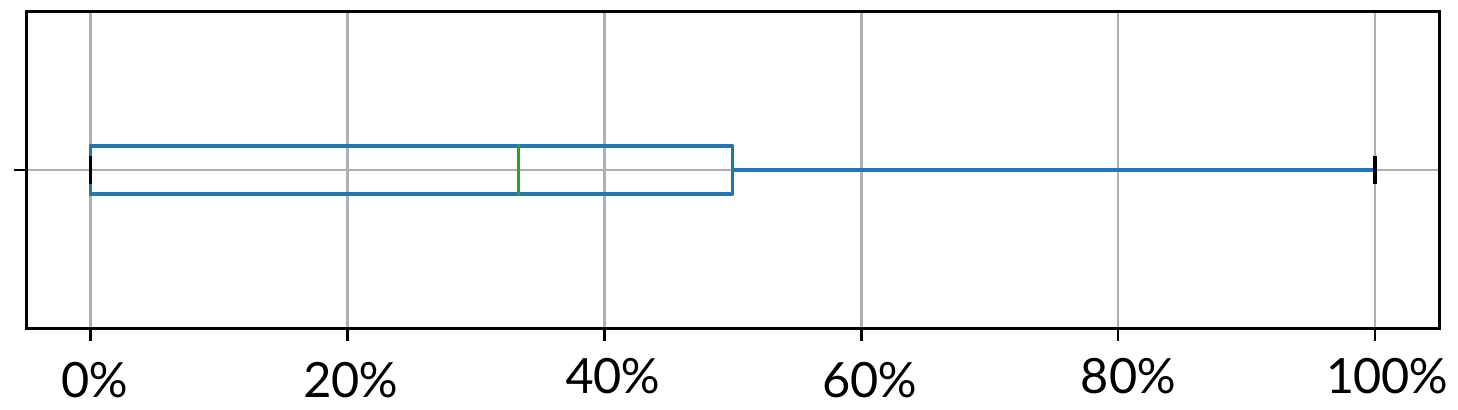}
	\caption{Percentage of non-specific S2Rs by bug report.}
	\label{fig:RQ5_results}
	\vspace{-1.5em}
\end{figure}

When the environment was reported, the information could be directly mapped into actions for reproducing the failure. That is, it was possible to select the right app version, Android version, and device for reproducing the failure.
In the case of OB, we had to look at the bug report discussion of six reports to better understand the problem associated with the reported failures, meaning that, in our analysis, the OB described in those bug reports was not specific enough for reproducing the failures.
Considering S2Rs, 73.9\% of the bug reports had at least one reported S2Rs that could not be directly mapped into a single GUI action but, instead, required multiple GUI actions. Based on the terminology defined in Section~\ref{sec:background}, this means that those bug reports had at least one non-specific S2R. Figure~\ref{fig:RQ5_results} reports the percentage of non-specific S2Rs in each bug report of our dataset. Across all reports, the S2Rs section had an average of 36\% of S2Rs that were non-specific. This results shows that there is the need to fill a gap to map S2Rs into corresponding GUI actions when reproducing reports.

Considering failure types, bug reports describing navigation failures had the highest average percentage of non-specific S2Rs (40\%), while output failures had the lowest (34\%). This result shows a minor difference in the specificity of S2Rs between reported failure types. There was also little difference in the average percentage of non-specific S2Rs reported by users (34.6\%) and developers (35.8\%).

\begin{tcolorbox}[colback=white, boxsep=0pt, left=4pt, right=4pt, before skip=4pt,after skip=0pt]
\textbf{RQ$_5$ answer:} Environment and OB information was specific enough to reproduce reported failures in the great majority of cases. A large percentage of reports (73.9\%) had at least one non-specific S2R, and the average percentage of non-specific S2Rs across all reports was 36\%. 
\end{tcolorbox}

\section{Discussion and Implications}
\label{sec:discussion}

\noindent\textbf{\textit{1) New automated techniques are needed for understanding non-crashing oracles.}} Most existing automated bug reproduction approaches for mobile apps focus on reproducing bugs leading to a crash~\cite{Fazzini2018,Zhao2019}. This is likely because failures related to crashes are easier to recognize, for example through detection of a crash dialog, and thus detect when a a crashing bug has been reproduced. 
However, our analysis shows that more than 70\% of the bug reports describe failures \textit{other than crashes} and thus require more sophisticated oracle definitions and detection.
For example, automated techniques for bug report reproduction might benefit from techniques that can define visual oracles using computer vision, such as detecting an incorrect color theme through color histogram analysis. Similarly, navigation failures might require analysis of statically computed program state graphs, to determine feasibility of navigation paths. Extending recent work on defining oracles through the derivation of program invariants (\eg~\cite{Su2021}) could further aid in oracle construction.

\noindent\textbf{\textit{2) There is a need for automated multi-modal understanding of bug report information.}} Our analysis has illustrated that bug reports can mix multiple modalities of information together in form of text, images, and recordings, which capture disparate pieces of information about a given bug. However, most recent work on automated bug report reproduction and analysis only considers the textual modality~\cite{Chaparro2019,Fazzini2018,Zhao2019}. Given the amount of prevalence of missing information, even in reproducible reports, revealed through our analysis of RQ$_3$, automated report analysis should strive to analyze \textit{all} types of reported information for a more robust and complete analysis. As such, new techniques for multi-modal understanding of bugs is needed. For example, deep learning techniques that connect images and natural language (e.g., dense image captioning~\cite{Johnson2016}) could be used to link textual information to visual information for more complete report analysis. Furthermore, in the case of S2Rs, automated techniques would also need to identify how to suitably order the information and this could be achieved by leveraging window transition graphs computed statically or dynamically from the apps~\cite{2019_ase_duling_goal,yang-jase18}.

\vspace{0.1in}
\noindent\textbf{\textit{3) Techniques for inferring and mocking app environments are essential.}} Historically, Android app developers have struggled to reign-in issues related to the fragmented platform and device ecosystem. These issues also surface in bug reporting. As identified while analyzing the bug reports considered, it is possible for bugs to manifest under specific combinations of device and platform versions. Considering, that this information is not always present in submitted bug reports (missing in 30\% of the cases), techniques that are able to infer, prioritize environmental settings (e.g., device and platform versions) are needed to help drive research on more advanced automated mobile bug report analysis techniques. Furthermore, considering that apps are released frequently~\cite{2019_ieccs_gao_on,mcilroy2016fresh} and bug reports do not always contain the associated app version, it would be beneficial to automatically infer the version of the app associated with a bug report. This task could be done by automatically by mapping bug report information into GUI components or code entities in the app.

\noindent\textbf{\textit{3) Reasoning about missing S2Rs is required.}} Our analysis illustrated that a large majority (92\%) of our studied bug reports have at least one missing S2R. This represents a notable challenge for automated report analysis techniques which will likely need to infer this missing information in order to provide robust analyses. Current techniques do offer advanced solutions (\ie they are based on random exploration) to help fill in certain missing gaps~\cite{Fazzini2018,Zhao2019,Chaparro2019}. However, additional techniques are likely needed that allow for fine-grained inference of missing steps. For instance, future techniques could examine existing corpora of bug reports (such the artifacts associated with this research) and attempt to infer missing steps via patterns learned from a corpus of complete bug reports.

\noindent\textbf{\textit{4) Handling non-specific S2Rs in bug report data is a major challenge.}} In addition to a high prevalence of missing S2Rs, our analysis also revealed that ~36\% of the S2Rs were mapped to multiple GUI actions. These S2Rs identified ``high-level'' operations, in which the actions or target GUI elements were not explicitly delineated. This situation represents a challenging reasoning problem for automated reproduction and report analysis techniques. Current techniques attempt to overcome such ambiguities through the use of ontological matching~\cite{Fazzini2018} or neural representations of text~\cite{Chaparro2019} in addition to random exploration. However, additional techniques for performing mapping of non-specific actions or targets are likely needed. For example future techniques may benefit from inferring descriptions of app controls or functionality through multi-modal image captioning models that allow for better mapping of text to runtime app information. Automated ``repair" of ambiguous bug report steps based on patterns learned form well-formed sets of reproduction steps may also be a worthwhile direction of exploration. Additionally, S2R descriptions could be extracted from sequences of GUI actions in existing test cases and be mapped to S2Rs in bug reports to facilitate their reproduction.

In summary, the analysis performed in this paper has revealed several notable implications that impact future work on automated bug report reproduction, reporting, analysis, and management. We believe that future work will benefit from these findings and the potential new directions of research that they point towards.
\section{Threats to Validity}

While we follow a systematic methodology in collecting, analyzing, and reporting our results, it is important to discuss the threats to validity of our study to provide a comprehensive view of our findings. In terms of external validity, our results may not generalize to bugs for other Android apps. However, given the number, diversity, and popularity of our subject applications and reports, we believe our studied reports should be reasonably representative of Android bug reports as a whole. We considered the most recent dataset of reproducible bug reports (with non-crashing bugs) and extended the dataset to also include developer-submitted bug reports. This dataset includes apps that vary in terms of their size and category. An additional threat could be posed by the fact that we only used open source apps. However, the evaluation includes apps such as
\textsc{Firefox Focus} and \textsc{Simplenote}, which have complex functionality and millions of installs. In terms of construct validity, our results might be affected by errors in the tools we used to perform our analyses. To mitigate this threat, we extensively tested our tools and multiple authors manually inspected the results. Finally, we also performed qualitative analyses, which could be impacted by divergent understanding among evaluators. To mitigate this threat, we used open coding based on negotiated agreement~\cite{2013_campbell_coding}.

\section{Related Work}
\subsection{General Studies on Bug Reports}
Related work investigated bug report properties to better understand multiple activities characterizing the bug report management process~\cite{2008_fse_bettenburg_what,Bettenburg2008duplicate,Sahoo2010,Breu2010,Davies2014,Ko2006,Ko2010-2,Thung2012,Guo2010}. Among different topics, this line of research analyzed bug report content, developers' and users' participation in bug report discussions, triaging, and bug fixing.
A prominent study carried out by Bettenburg et al.~\cite{2008_fse_bettenburg_what} identified desired aspects that should be contained in a bug report. In follow-up work, Bettenburg et al.~\cite{Bettenburg2008duplicate} also showed that duplicated bug reports contain some additional helpful information that could be used for bug triaging. Sahoo et al.~\cite{Sahoo2010} identified the main components necessary for bug reproduction by performing an empirical study.
Some prior studies focused primarily on user-submitted bug reports. This line of research investigated how users typically communicate software problems~\cite{Ko2006}, the usefulness of the provided information by power users~\cite{Ko2010-2}, and user communitys' expectations~\cite{Chilana2010}.
In this paper, we investigated key aspects related to both user-submitted and developer-submitted Android bug reports. Furthermore, we focused on the aspects related to the reproduction of bug reports and specifically investigated how the bug report information relates to the information needed to reproduce the reports.

\subsection{Bug Report Studies for Mobile Apps}
Most of the initial studies on bug reports focused on desktop applications. However, because of smartphone apps' availability, usability, and popularity in the last decade, researchers have also started focusing on studying characteristics of bug reports for mobile apps. Zhou et al.~\cite{Zhou2015} performed a study to understand the bug management between desktop and mobile software. Bhattacharya et al.~\cite{Bhattacharya2013} studied mobile bug reports and the bug-fixing process. Aljedaani et al.~\cite{Aljedaani2019} compared the bug reports between Android and iOS. Zhang et al.~\cite{Zhang2019LabellingIR} studied mobile apps bug reports, labeled those reports, and computed similarities with the previously labeled ones. In our study we reproduced bug reports, characterized the failures associated with the reports, analyzed the usefulness of the information provided in the reports, and categorized the reporting modalities. Previous studies also produced datasets of Android bugs with associated bug reports. Wendland et al.~\cite{Wendland2021} created a dataset of reproducible, user-submitted bug reports. Su et al.~\cite{2021_fse_benchmarking_su} created a dataset of crashing bugs based on GitHub issues. Fazzini et al.~\cite{Fazzini2018} and Zhao et al.~\cite{Zhao2019} also assembled a dataset of crashing bugs for their research on automated reproduction of bug reports. Compared to these datasets, to the best of our knowledge, this paper is the first to create and consider in its study a dataset of non-crashing and reproducible bug reports that contains both user-submitted and developer-submitted reports.
\section{Conclusion}

We presented an empirical study that characterized reproducible Android bug reports. Specifically, we manually reproduced 180 bug reports systematically mined from Android apps on GitHub and investigated how the information contained in the bug report relates to the task of reproducing the reports. Our analysis identified that reported failures can be grouped into four categories, three of which are not yet considered by existing automated reproduction techniques, reporters use different modalities to report the information relevant for reproducing failures, a large number of reports (74\%) have at least one non-specific S2R (\ie multiple GUI action are necessary to perform the operation described by the S2R), the great majority of reports (92\%) do not provide all the S2Rs that are necessary to reproduce the reports, and  bug report discussions can, in some cases (19\%), provide additional information useful for the reproduction of the reports.

In future work, we first plan to present our findings to Android developers and then develop techniques to aid automated reproduction of bug reports. To support automated reproduction of bug reports, we first plan to define an approach that leverages natural language processing and computer vision techniques to automatically encode OB information into oracles and so aid reproduction of output, cosmetic, and navigation failures. Second, we plan to define a technique that combines S2Rs information reported using different modalities. Third, we plan to define a technique that leverages the information contained in existing test cases to help mapping non-specific S2Rs to corresponding GUI actions. Finally, we believe that additional studies into the reproduction of bug reports for software in other domains are needed and those studies could inform techniques for bug report management in those domains.

\vspace{-1em}
\section*{Acknowledgment}
\vspace{-1em}
This work was partially supported by a gift from Facebook and the NSF CCF-2007246 \& CCF-1955853 grants. Any opinions, findings, and conclusions expressed herein are the authors' and do not necessarily reflect those of the sponsors.

\balance

\bibliographystyle{IEEEtran}
\bibliography{IEEEabrv,paper}

\end{document}